# Allocation de ressources dans un réseau de radio cognitive en utilisant JADE

Benmammar Badr

# RÉSUMÉ

Dans ce rapport, nous nous intéressons à la négociation entre plusieurs PU et plusieurs SU dans un contexte radio cognitive en utilisant les systèmes multi-agents. Pour cela, nous avons implémenté des coalitions de PU (CPU) et de SU (CSU). Nous avons également utilisé la décision multicritère (algorithme TOPSIS) pour répondre au mieux aux besoins des SU.

La gestion par coalitions ainsi que la technique d'agrégation que nous avons utilisée dans ce contexte réduisent considérablement le nombre de messages échangés. Le traitement des négociations de manière simultanée par plusieurs CSU minimise le temps de réponse côté SU.

**Mots de clés :** Gestion de spectre, Radio Cognitive, Système multi-agents, JADE, Négociation, Agrégation, Coalition, Décision multicritère.

# ABSTRACT

In this work, we are interested in the negotiation between several PU and several SU in a cognitive radio context by using multi-agent systems. So, we have implemented PU coalitions (CPU) and SU coalitions (CSU). We also used the multi-criteria decision (TOPSIS algorithm) to best meet the needs of SU.

Management by coalitions and the aggregation technique that we have used in this context significantly reduce the number of messages exchanged in the network. Processing negotiations simultaneously by several CSU minimizes response time side SU.

# TABLE DES MATIÈRES

# INTRODUCTION GÉNÉRALE

# Introduction générale

Durant cette dernière décennie, les télécommunications ont pris un essor considérable dans les domaines techniques et économiques répondant ainsi aux exigences de nos sociétés actuelles.

La croissance des services sans-fil démontre la demande incessante pour les communications sans-fil, bien que le spectre soit de plus en plus encombré. Cependant des études établies par la Commission Fédérale de Communication (CFC) en 2003 ont montré que la plupart du spectre est sous exploitée.

Pour éviter l'encombrement et la sous-utilisation du spectre, on a fait recours à la nouvelle technologie dite Radio Cognitive (RC). Celle-ci permet à des usagers sans Licence (appelés: usagers secondaires « SU ») d'utiliser des fragments du spectre sans nuire aux communications des usagers avec licence (appelés: usagers primaires « PU »). Ces derniers, propriétaires de licence d'exploitation, sont toujours prioritaires à l'accès aux ressources spectrales.

A titre d'information, la notion de la RC a été introduite pour la première fois par Mitola et Maguire [13] en 1999.

L'objectif de notre travail est de réaliser une négociation entre plusieurs PU et plusieurs SU en utilisant les systèmes multi-agents pour l'accès dynamique au spectre dans les réseaux de radio cognitive.

Pour cela, on a implémenté des agents coalitions CPU et CSU de la façon suivante:

Chaque CPU est un coordinateur d'un nombre limité de PU se trouvant dans sa zone géographique pendant les négociations avec les CSU. Ces derniers regroupent et représentent des SU avec des tailles non nécessairement homogènes tout en respectant la localisation imposée.

L'utilisation des agents coalitions (CPU et CSU) avec la technique d'agrégation de messages pendant la négociation contribuent à réduire le temps de réponse et le nombre de messages échangés dans le réseau.

Notre rapport est organisé comme suit :

**Le chapitre I** est consacré à l'étude de la radio cognitive (sa définition, son fonctionnement, son architecture ainsi que ses différents domaines d'applications), aux systèmes multi-agents (modèle d'interaction entre les agents, les plateformes de SMA en particulier JADE) et à l'introduction de quelques méthodes de décision multicritères, notamment la méthode TOPSIS.

# Introduction générale

**Le chapitre II** est destiné à la présentation de notre application de gestion dynamique de spectre. Cette application est réalisée à l'aide du simulateur JADE. Une discussion par rapport aux résultats obtenus est aussi présente dans ce chapitre.

# CHAPITRE I : Radio cognitive, systèmes multi-agents et décisions multicritères

## I.1  Introduction

Ce chapitre est consacré à l'étude de la radio cognitive, son fonctionnement, son architecture ainsi que ses différents domaines d'applications. Nous présentons aussi la notion de systèmes multi-agents, leurs plateformes et leurs modèles d'interactions. Enfin, nous passons en revue quelques méthodes de décision multicritère en particulier celle de TOPSIS.

## I.2  Radio cognitive

### I.2.1  Définition de la radio cognitive

La Radio Cognitive RC [26] est un paradigme pour les communications sans-fil dans lequel un réseau ou un nœud est capable de modifier de manière automatique ses paramètres de transmission ou de réception afin de communiquer efficacement tout en évitant les interférences avec d'autres utilisateurs agréés ou non. Cette auto-configuration et auto-adaptation des paramètres sont basées sur une surveillance active de plusieurs facteurs dans l'environnement interne ou externe à la radio tels que le spectre de la Radio Fréquence RF, le comportement de l'utilisateur et l'état du réseau. L'idée de radio cognitive se développe autour du concept plus ancien de radio logicielle.

### I.2.2  Réseau cognitif

Un réseau cognitif coordonne les transmissions suivant différentes bandes de fréquences et différentes technologies en exploitant les bandes/technologies disponibles à un instant donné et à un endroit fixé.

Il a besoin d'une station de base capable de travailler sur une large gamme de fréquences afin de reconnaître différents signaux présents dans le réseau et se reconfigurer intelligemment.

- ❖ **Exemple** [11]**:** les stations de base détecteront la présence de différentes technologies, et se reconfigureront (changeant d'une station GSM à une station UMTS si des terminaux UMTS sont présents) afin de s'adapter à des standards ou des services à fournir à un instant précis.

### I.2.3  Architecture de radio cognitive

L'architecture de RC [5] est un ensemble cohérent de règles de conception par laquelle un ensemble spécifique de composants réalise une série de fonctions.

Dans cette partie, nous allons traiter l'architecture d'une radio cognitive, dans laquelle des SDR (Software Defined Radio), des capteurs… seront intégrés pour créer une meilleure qualité d'information (QoI) et développer des capacités à observer, orienter, planifier, décider, agir et apprendre l'environnement de l'utilisateur et les ressources fréquentielles.

Un nœud AACR (Adaptative Aware Cognitive Radio) est composé d'une série minimaliste de 6 composantes fonctionnelles. Un élément fonctionnel est une boîte noire à laquelle des fonctions ont été attribuées, mais pour laquelle la mise en œuvre des composants n'est pas précisée. Ainsi, bien que les applications de ces composants soient susceptibles d'être principalement de type logiciel, les détails de ces composants logiciels ne sont pas précisés.

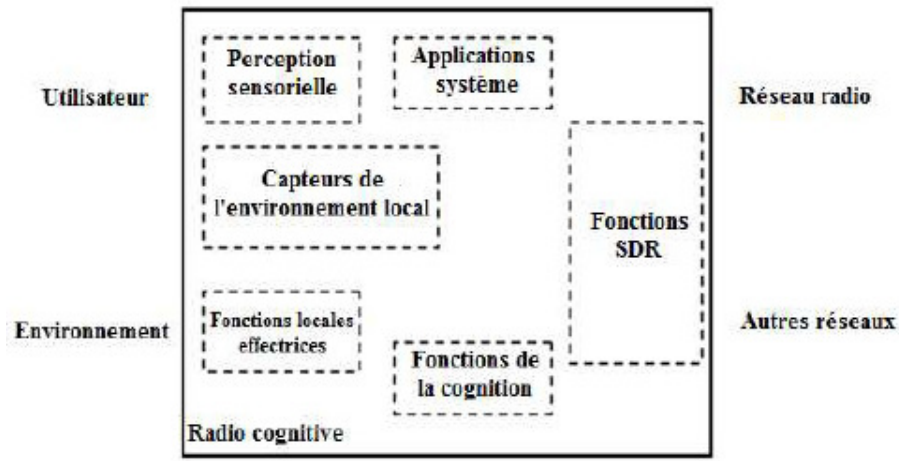

**Figure I.1:** Architecture de la radio cognitive [8]

Les six composantes fonctionnelles sont:

1. La perception sensorielle (SP) de l'utilisateur incluant l'interface haptique (du toucher), acoustique, la vidéo, les fonctions de la détection et de la perception. Les fonctions SP de l'utilisateur peuvent inclure un matériel optimisé, par exemple, pour calculer les vecteurs de flux vidéo en temps réel pour aider la perception d'une scène.
2. Les capteurs de l'environnement local (emplacement, la température, l'accéléromètre, compas, etc.)
3. Les applications systèmes (les services médias indépendants comme un jeu en réseau).
4. Les fonctions SDR (incluant la détection RF et les applications radio de la SDR).

5. Les fonctions de la cognition (pour les systèmes de contrôle, de planification et de l'apprentissage).
6. Les fonctions locales effectrices (synthèse de la parole, du texte, des graphiques et des affiches multimédias).

Généralement, les communications vocales avec un annuaire téléphonique, la messagerie texte et l'échange de photos ou de clips vidéo correspondent aux principales applications pour les systèmes SDR. Les applications de l'AACR vont au-delà des services offerts par les systèmes SDR pour une plus grande flexibilité personnelle et le choix de la connectivité sans-fil. L'utilisateur pourrait contrôler le passage de l'AACR d'un réseau à un autre en fonction du coût de la connexion.

### I.2.4 Le cycle de la cognition

Le cycle de la cognition [14] est un schéma global décrivant l'ensemble des interactions entre les différents modules du système y compris l'environnement externe. L'analyse ou l'observation de son environnement, la planification de ses actions et la décision constituent le fondement de ses activités tel que présenté dans la figure I.2

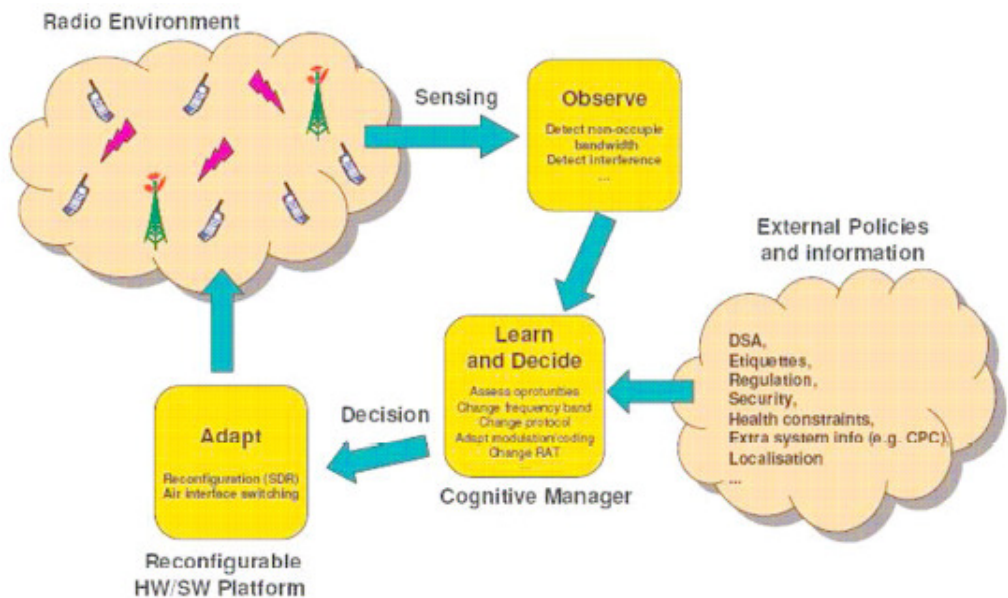

**Figure I.2:** Vision de Mitola (Cycle cognitif) [11]

La radio cognitive [1], suivant le cycle de la cognition, atteint le niveau d'autonomie espéré et une capacité de configuration dynamique de ses paramètres. En effet, l'observation (Observe) de son environnement caractérisant la collecte d'informations lui assure la prise de décision (Decide) la mieux adaptée. Ce processus transite par une suite d'étapes entre autres, l'orientation (Orient) qui est primordiale dans l'établissement des priorités, la classification par affinité ou en fonction du besoin ainsi que la planification (Plan) des actions possibles, l'apprentissage (learn) motivé par l'expérience acquise grâce à l'environnement.

### I.2.5 Fonctionnement de la radio cognitive

Les fonctions suivantes sont principales et indispensables pour le bon fonctionnement de la radio cognitive.

Les bandes spectrales inutilisées ont des caractéristiques différentes les unes des autres. Ces caractéristiques sont la fréquence d'opération de la bande spectrale, le débit et le temps. Toutes ces informations changent au cours du temps vu la nature dynamique de l'environnement radio. C'est dans ce contexte que les auteurs [6] ont présenté les nouvelles fonctions requises pour gérer les ressources spectrales dans les réseaux de radio cognitive.

a. Gestion de spectre:
- **L'analyse spectrale:** elle permet de caractériser les différentes bandes spectrales en termes de fréquence d'opération, de débit, de temps et de l'activité de l'Utilisateur Primaire (UP). Cette caractérisation sert à répondre aux exigences de l'URC (Utilisateur à Radio Cognitive). Des paramètres supplémentaires viennent compléter cette caractérisation, à savoir, le niveau d'interférence, le taux d'erreur du canal, les évanouissements, le délai et le temps d'occupation de la bande spectrale par un URC [7].
- **Décision spectrale:** après que toutes les bandes spectrales aient été catégorisées et c1assifiées, on applique un ensemble de règles décisionnelles pour obtenir la ou les bandes spectrales les plus appropriées à la transmission en cours, en tenant compte des exigences de l'URC [10].

b. Mobilité de spectre:

C'est le processus qui permet à l'utilisateur de la radio cognitive de changer sa fréquence de fonctionnement.

Les réseaux de radio cognitive essaient d'utiliser le spectre de manière dynamique en permettant à des terminaux radio de fonctionner dans la meilleure bande de fréquence disponible, de maintenir les exigences de communication transparentes au cours de la transition à une meilleure fréquence [8].

    c. Détection de spectre:

Cette phase a pour objectif la détection de l'état de spectre (libre ou utilisé), afin de le partager avec son utilisateur primaire tout en évitant toutes interférences. Le défi réside dans le fait de mesurer l'interférence au niveau du récepteur primaire causée par les transmissions d'utilisateurs secondaires [4].

### I.2.6 Avantages de la radio cognitive

La radio cognitive apportera des solutions aux problèmes liés aux allocations fixes des fréquences. Elles se trouvent alors toutes utilisées, ce qui contribuera à plus de souplesse dans l'architecture et dans la conception des émetteurs et des récepteurs.

La radio cognitive, par sa capacité à modifier au cours d'une communication ses caractéristiques: fréquence, puissances du signal, contribuera à améliorer les performances globales du système en assurant en permanence des performances optimales en fonction de l'entourage et de la position.

La radio cognitive facilitera aussi l'interopérabilité entre les différents réseaux et standards de télécommunication actuels.

Remarquons cependant que ces différents avantages induisent bien évidemment des émetteurs/récepteurs qui seront à la fois un peu plus complexes et difficiles à réaliser [15].

### I.2.7 Accès dynamique au spectre dans le contexte de la RC

Nous avons noté au cours de ces dernières années une croissance fulgurante des services sans-fil, et ce, d'après un flux de demandes de communication, entrainant un important encombrement au niveau du spectre. Ce n'est un secret pour personne, l'allocation du spectre statique est un problème qui sème le trouble dans les réseaux sans-fil. Communément, ces allocations engendrent un usage inopérant et inefficace du spectre en formant ce que l'on nomme trous ou espaces blancs dans le spectre. Pour résoudre le problème de l'encombrement, les RRC se servent de l'accès dynamique au spectre et font appel à des techniques d'accès utiles et efficaces afin de parvenir au spectre requis quand et où nécessaire (si possible), à un coût accessible.

a. **Accès au spectre en utilisant les enchères:**

Les enchères s'appuient sur le concept de vente et d'achat des biens ou de services. La visée majeure de l'usage de ces enchères dans les RRC est l'apport d'une motivation aux utilisateurs secondaires de manière à optimiser pleinement l'utilisation du spectre.

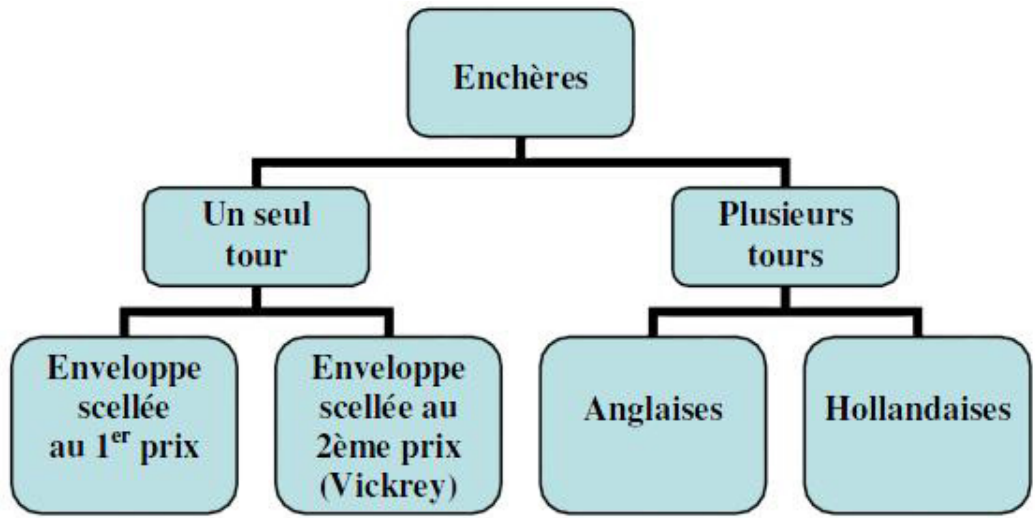

**Figure I.3:** Organigramme représentant les types d'enchères [16]

b. **Accès au spectre en utilisant la théorie des jeux:**

La théorie des jeux a connu un essor considérable dans la recherche scientifique durant la dernière décennie, grâce a son efficacité et son exactitude dans la modélisation des comportements des individus, et la prévision des résultats en fonction des conditions initiales et des décisions prises. Il y a deux types: jeux coopératifs et jeux compétitifs.

c. **Accès au spectre en utilisant les chaînes de Markov:**

Les approches de la théorie des jeux ne conçoivent pas de modèle théorique pour ce qui concerne l'interaction entre les utilisateurs secondaires et primaires pour l'abord au spectre. Cette modélisation peut être accomplie en se servant parfaitement des chaînes de Markov.

d. Accès au spectre en utilisant les Systèmes Multi-agents (SMA).

## I.3 Les Systèmes Multi-Agents (SMA)

### I.3.1 Historique sur les méthodes de programmation

Nous proposons un bref aperçu des différentes programmations utilisées à ce jour. Dans ce travail, on s'intéresse à la programmation orientée agents.

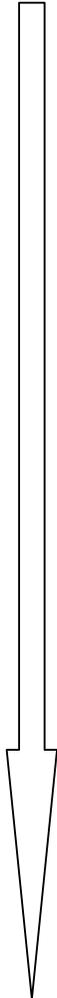

| Langage Machine | Chaque famille de CPU possède son porophore jeu d'instructions |
|---|---|
| **Assembleur** | Langage bas niveau |
| **Programmation Procédurale** | Sous programmes: Procédures, fonctions (Basic, Pascal, C, Fortran,..) |
| **Programmation Orientée Objet** | Objet = Etat+ Comportement + Identité Concepts fondamentaux : Objet, classe, héritage, polymorphisme, encapsulation (C++, JAVA, C#, ..) |
| **Programmation Orientée Objet Distribués** | Objets distribués sur plusieurs machines Middlewares : (RMI, CORBA, JMS, …) |
| **Programmation Orientée Composants** | Objets distribués, réutilisables, configurables, Interchangeables, évolutifs, mobiles, surveillable à chaud : Conteneur (EJB) |
| **Programmation Orientée Services** | Composant disponibles à d'autres applications distantes hétérogènes via des protocoles (http) transportant des données: XML, JSON => SOAP et REST |
| **Programmation Orientée Agents** | Service + Intelligence + Apprentissage+… |

**Tableau I.1:** Evolution de la programmation [9]

### I.3.2 Définition d'agent

Un agent peut être un processus, un robot, un être humain, etc…, la notion d'agent est utilisée dans beaucoup de domaines: sociologie, biologie, psychologie cognitive, psychologie sociale, informatique [17]. Dans ce dernier domaine, les auteurs dans [33] par exemple ont utilisé les agents pour la négociation de la QoS entre un fournisseur de services et un utilisateur final.

Un Agent en informatique est une entité réelle ou abstraite qui fonctionne continuellement et de manière autonome dans un environnement où d'autres processus se déroulent et d'autres agents existent pour atteindre les objectifs pour lesquels il a été conçu. Dans un SMA, un agent peut communiquer directement avec d'autres agents et doter de capacités semblables aux êtres vivants.

### I.3.3 Propriétés d'un Agent

Les chercheurs en intelligence artificielle s'accordent sur la nécessité de l'existence de quelques propriétés pour qu'on puisse parler d'agents intelligents. Voici les cinq propriétés d'un agent intelligent [12]:

1. **Autonome :** agir sans l'intervention directe d'un humain (ou d'un autre agent) en contrôlant ses actions et son état interne.

2. **Proactif :** capable d'avoir un comportement opportuniste, dirigé par ses buts ou sa fonction d'utilité, et prendre des initiatives au moment approprié.

3. **Flexible :** capable de répondre à temps

4. **Social :** capable d'interagir avec les autres agents (artificiels ou humains)

5. **Situé :** recevoir des entrées sensorielles provenant de son environnement et ainsi effectuer des actions qui sont susceptibles de changer cet environnement.

### I.3.4 Dichotomie entre les agents réactif et cognitif

#### a. Agents réactifs

Agents sans intelligence (sans anticipation, sans planification) qui réagissent par stimulus-réponse à l'état courant de l'environnement. Des comportements intelligents peuvent émerger de leur association.

- ✓ Pas de représentation explicite.
- ✓ Organisation implicite/induite.
- ✓ Communication via l'environnement.

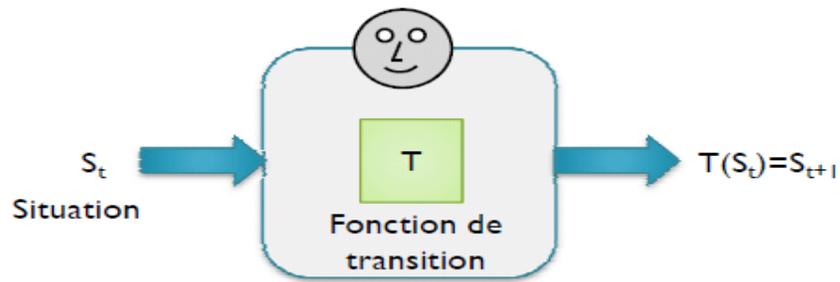

**Figure I.4:** Fonction d'un Agent réactif [9]

**b. Agents cognitifs**

Chaque agent est spécialisé dans un domaine et sait communiquer avec les autres. Ils possèdent des buts et des plans explicites leur permettant de les accomplir. (analogie avec les groupes en sociologie).

- ✓ Représentation explicite de soi, environnement et les autres agents.
- ✓ Organisation explicite.
- ✓ Interaction explicite et élaborée.

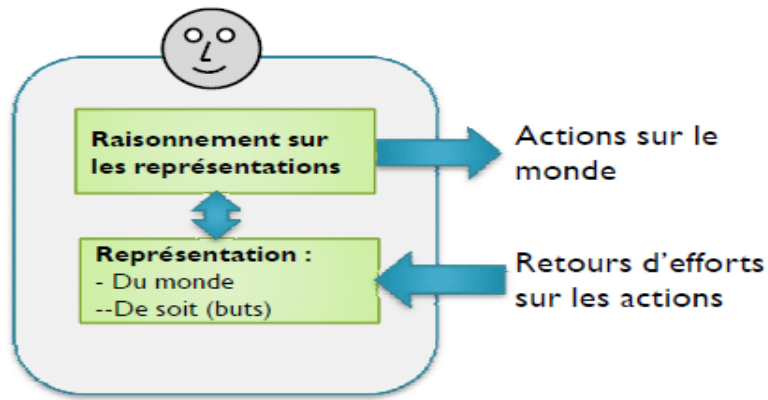

**Figure I.5:** Fonction d'un Agent cognitif [9]

## I.3.5 Définition de SMA

D'après Anne Nicole (Laboratoire Intelligence Artificielle - Caen), un SMA comporte plusieurs agents qui interagissent entre eux dans un environnement commun. Certains de ces agents peuvent être des personnes ou leurs représentants (avatars), ou même des machines mécaniques. S'il y a moins de trois agents, on parle plutôt d'interaction homme/machine, ou machine/machine que de systèmes multi-agents.

### I.3.6 Utilisation d'un SMA

L'utilisation d'un SMA se fait lorsque le problème est trop complexe mais décomposable et parfois pour gagner du temps (paralléliser le problème). Aussi quand le problème proposé n'a pas de solution générale ou qu'elle est trop coûteuse en CPU.

A des fins de modélisation (populations, structures moléculaires, réseaux de spins, tas de sables...) et pour avoir une certaine robustesse (redondance); on applique un SMA.

### I.3.7 Applications des SMA

Les SMA sont à l'intersection de plusieurs domaines scientifiques: informatique répartie et génie logiciel, intelligence artificielle, vie artificielle. Ils s'inspirent également d'études issues d'autres disciplines connexes notamment la sociologie, la psychologie sociale, les sciences cognitives et bien d'autres. C'est ainsi qu'on les trouve parfois à la base des:

- ✓ Systèmes distribués.
- ✓ Interface personnes-machines.
- ✓ Bases de données et bases de connaissances distribuées coopératives.
- ✓ Systèmes pour la compréhension du langage naturel.
- ✓ Protocoles de communication et réseaux de télécommunications.
- ✓ Programmation orientée agents et génie logiciel.
- ✓ Robotique cognitive et coopération entre robots.
- ✓ Applications distribuées comme le web, l'Internet, le contrôle de trafic routier, le contrôle aérien, les réseaux d'énergie, etc.

### I.3.8 Interaction entre agents

Toute organisation dispose d'une interaction comme composant de base qui est à la fois source et produit de la constance et durabilité de cette organisation, et la dissolution d'une organisation coïncide à la disparition (ou en tout cas à la diminution) des interactions des personnes présentes dans cette organisation.

Une des propriétés essentielles de l'agent dans un SMA est celle d'interagir avec les autres agents.

Ces interactions sont souvent expliquées comme étant tout type d'action réalisée au sein du système d'agents et qui a la possibilité d'apporter une modification du comportement de l'autre agent.

Les types d'interaction:
- **La négociation** est un processus de communication d'un groupe d'agents permettant d'atteindre un accord mutuellement accepté.

  Les grands types de négociation :
  - ✓ **Négociation compétitive:** les agents d'intérêts différents tentent un choix de groupe sur des alternatives bien définies.
  - ✓ **Négociation coopérative:** les agents ont un but commun aussi les agents sont collaboratifs, ils coopèrent.

- **La coopération** est bel et bien d'une parfaite action collective dont il est question, les agents travaillent en groupe et s'affèrent à la résolution d'une difficulté commune. Leur mission est d'optimiser et maximiser leur propre satisfaction et participer à la réussite d'une entité**.**

- **La coordination** est une question de taille pour les SMA et la résolution de systèmes distribués. Elle est centrale car sans elle, un important problème est à prévoir. Le groupe d'agents se verra dégénérer et sombrer dans le chaos. Un groupe d'individus qui ne jouit d'aucune règle.

### I.3.9   Les plateformes SMA

Il existe des plateformes qui permettent la prise en charge des fonctions de base d'un simulateur multi-agents comme la communication, le cycle de vie des agents, la perception et l'environnement.

Parmi les plateformes les plus connus il y a Madkit [27], Mason [28], NetLogo [25] et Gama [18]. Cependant, ces plateformes n'intègrent pas de mécanismes de parallélisme, il est nécessaire de développer une sur-couche à la main, pour distribuer ou paralléliser une simulation.

Un autre type de plateformes multi-agents prennent en compte nativement le parallélisme comme RepastHPC [21], D-Mason [23], Pandora [19], FLAME [26], JADE (Java Agent

DEvelopment) [20] est utilisée pour la réalisation de notre application sera présentée dans l'annexe de ce rapport.

## I.4 La décision multicritères

### I.4.1 Définition de la décision

On présente souvent la décision comme le fait d'un individu isolé exerçant librement un choix entre plusieurs possibilités d'actions.

### I.4.2 L'aide à la décision multicritère

Constitue une branche d'étude majeure de la recherche opérationnelle impliquant plusieurs écoles de pensée, principalement américaine avec les travaux de Thomas L. Saaty [31] et européenne avec ceux de Bernard Roy [22] du LAMSADE (Laboratoire d'analyse et modélisation de systèmes pour l'aide à la décision).

Il s'agit de méthodes et de calculs permettant de choisir la meilleure solution ou la solution optimale parmi tout un ensemble de solutions, l'alternative de type OUI-NON n'étant qu'un cas particulier du cas général.

### I.4.3 Domaines d'application

Les méthodes multicritères peuvent s'appliquer dans des contextes très différents: entreprises privées / administration, décision stratégique / décision répétitive et routinière. Parmi les exemples qui peuvent utiliser la décision multicritères, le choix d'un site d'aménagement et décision d'investissement. Ainsi le choix de l'utilisation d'une technologie ou d'un système d'information.
Afin de choisir un candidat pour un poste ou sélection de fournisseurs, il doit appliquer une des méthodes de décision multicritères.

### I.4.4 Illustration des méthodes multicritères

Nous présentons dans ce qui suit les méthodes multicritères, en fonction de la problématique considérée. Les méthodes suivantes sont préconisées pour répondre à une problématique de choix.

**a. MÉTHODE WSM** (Weight Sum Method)

Également appelée la somme pondérée, cette méthode est la plus simple des méthodes multicritère. C'est aussi la méthode la plus utilisée dans la vie quotidienne.

Elle requiert que les critères soient quantitatifs, qu'ils aient tous la même unité et qu'ils s'étendent sur une même échelle ou gamme de valeurs ou qu'ils soient tous normalisés.

**b. MÉTHODE WPM** (Weight Product Method)

La multiplication de ratios se nomme en anglais Weight Product Method (WPM). Elle évite certains défauts de la somme pondérée [24] et s'en différencie principalement par les deux points suivants :

Les critères (qui doivent être quantitatifs) peuvent avoir des gammes de valeurs sur des échelles différentes les unes des autres. Chaque critère à sa propre échelle qui lui est adaptée. Ainsi, on conserve une certaine homogénéité de la prise en compte de tous les critères.

**c. MÉTHODE AHP** (Analytic Hierarchy Process) [3] a été proposée par Thomas L. Saaty en 1971 pour modéliser les processus de prise de décisions subjectives basées sur des critères multiples dans un système hiérarchique. Cette méthode est très pratique pour déterminer les poids relatifs à des critères. Trois des méthodes les plus utilisées pour déterminer les pondérations dans AHP sont:

- ✓ Moyenne des colonnes normalisées (ANC).
- ✓ Normalisation de la moyenne de la ligne (NRA).
- ✓ Normalisation de la moyenne géométrique des lignes (NGM).

**d. MÉTHODE ELECTRE** (ELimination Et Choix Traduisant la REalité) est une famille de méthodes d'analyse multicritères développée en Europe.

La méthode ELECTRE I a été élaborée par Bernard Roy en 1968, avec l'aide de P. BERTIER. Il a ensuite développé la méthode ELECTRE II (B. Roy, P. Bertier 1971) [30], comme réponse aux méthodes de prise de décision existantes. Cette méthode pourrait être considérée comme une philosophie d'une aide à la décision.

**e. MÉTHODE TOPSIS**

La méthode « TOPSIS » (Technique for Order Preference by Similarity to Ideal Solution) a été implémentée dans le cadre de notre travail.

#### e.1. Définition et origine

TOPSIS est une méthode d'analyse multicritère pour l'aide à la prise de décision. Elle a été introduite par Yoon et Hwang en 1981 [29].

#### e.2. Idée principale de TOPSIS

Selon ce procédé, la solution la plus adaptée serait la distance la plus courte à partir de la solution idéale et la plus grande distance de la solution anti-idéale.

#### e.3. Solutions Idéale et Anti-Idéale

- Idéale Solution:  $A^* = \{ g_1^*, ..., g_j^*, ..., g_n^* \}$

    avec $g_j^*$ la meilleure valeur pour le $j^{ème}$ critère parmi toutes les actions.

- Anti-Idéale Solution $A' = \{g_1', ..., g_j', ..., g_n' \}$.

    avec $g_j'$ la plus mauvaise valeur pour le $j^{ème}$ critère parmi toutes les actions.

#### e.4. Matrice de décision

TOPSIS suppose que nous avons m alternatives (options) et n attributs / critères et nous avons le score de chaque option par rapport à chaque critère.

Soit $x_{ij}$ xijscore de l'option i par rapport au critère j.

Nous avons une matrice D= $(x_{ij})$ matrice de n×m.

Soit J l'ensemble des avantages ou attributs critères (plus est mieux).

Soit J' l'ensemble d'attributs ou de critères négatifs (moins est plus).

$$\mathbf{D} = \begin{matrix} & \begin{matrix} C_1 & C_2 & \cdots & C_n \end{matrix} \\ \begin{matrix} A_1 \\ A_2 \\ \vdots \\ A_m \end{matrix} & \begin{bmatrix} x_{11} & x_{12} & \cdots & x_{1n} \\ x_{21} & x_{22} & \cdots & x_{2n} \\ \vdots & \vdots & \ddots & \vdots \\ x_{m1} & x_{m2} & \cdots & x_{mn} \end{bmatrix} \end{matrix}$$

#### e.5. Algorithme de TOPSIS

Les six étapes de l'algorithme TOPSIS :

**Etape 1:** Calcul des préférences normalisées   (normalized ratings)

$r_{ij} = x_{ij} / (\Sigma x^2_{ij})$        i = 1… m     j = 1… n

**Etape 2:** Calcul des préférences normalisées avec des poids associés aux critères (weighted normalized ratings)

$v_{ij} = w_j \, r_{ij}$     $i = 1,\ldots, m$     $j = 1,\ldots, n$

$$V = \begin{bmatrix} v_{1_1} & \cdots & v_{1n} \\ \vdots & \ddots & \vdots \\ v_{m_1} & \cdots & v_{mn} \end{bmatrix} = \begin{bmatrix} w_1.r_{1_1} & \cdots & w_n.r_{1n} \\ \vdots & \ddots & \vdots \\ w_1.r_{m_1} & \cdots & w_n.r_{mn} \end{bmatrix}$$

**Etape 3:** Identification des solutions idéales et anti-idéales

$A^* = \{ v_1^*, \ldots, v_j^*, \ldots, v_n^* \}$ , où

$v_j^* = \{(\max (v_{ij})/j \in J_1), (\min (v_{ij}) / j \in J_2)\}$

$A' = \{ v_1', \ldots, v_j', \ldots, v_n' \}$, où

$V_j' = \{(\min (v_{ij})/j \in J_1), (\max (v_{ij}) / j \in J_2)\}$

$J_1$ : ensemble des critères de bénéfice.

$J_2$ : ensemble des critères de coût.

**Etape 4:** Calcul des distances (Separation measures)

$S_i^* = \left[\sum_j (V_j^* - V_{ij})^2\right]^{1/2}$     i=1… m

$S_i' = \left[\sum_j (V_j' - V_{ij})^2\right]^{1/2}$     i=1… m

**Etape 5:** Calcul de l'index de similarité à la solution idéale

$$C_i^* = S'_i / (S_i^* + S'_i) , \qquad 0 < C_i^* < 1$$

**Etape 6:** Ordre de préférence
- Choisir l'action ayant le plus grand index de similarité (problématique de choix).
- Ranger les actions par ordre décroissant des index de similarité (problématique de rangement).

### e.6. Exemple d'utilisation de la méthode TOPSIS [32]

| Poids $(w_j)$ | 0.1 | 0.4 | 0.3 | **0.2** |
|---|---|---|---|---|

- La matrice de décision:

|  | Style | Fiabilité | Carburant Eco. | Coût |
|---|---|---|---|---|
| Civic | 7 | 9 | 9 | 8 |
| Saturn | 8 | 7 | 8 | 7 |
| Ford | 9 | 6 | 8 | 9 |
| Mazda | 6 | 7 | 8 | 6 |

- **Etape 1:** Calculer $\sqrt{(\sum_i x_{ij}^2)}$ pour chaque colonne, puis diviser chaque $x_{ij}$ de la matrice de décision sur la valeur calculée de sa colonne pour obtenir $r_{ij}$

|  | Style | Fiabilité | Carburant Eco. | Coût |
|---|---|---|---|---|
| Civic | 0.46 | 0.61 | 0.54 | **0.53** |
| Saturn | 0.53 | 0.48 | 0.48 | 0.46 |
| Ford | 0.59 | 0.41 | 0.48 | 0.59 |
| Mazda | 0.40 | 0.48 | 0.48 | 0.40 |

- **Etape 2:** Multiplier chaque colonne par le poids associé $w_j$ pour avoir le $V_{ij}$

|  | Style | Fiabilité | Carburant Eco. | Coût |
|---|---|---|---|---|
| Civic | 0.046 | 0.244 | 0.162 | 0.106 |
| Saturn | 0.053 | 0.192 | 0.144 | 0.092 |
| Ford | 0.059 | 0.164 | 0.144 | 0.118 |
| Mazda | 0.040 | 0.192 | 0.144 | 0.080 |

- **Étape 3:** a- déterminer la solution idéale A*

    **A* = {0.059, 0.244, 0.162, 0.118}**

|        | Style  | Fiabilité | Carburant Eco. | Coût   |
|--------|--------|-----------|----------------|--------|
| Civic  | 0.046  | **0.244** | **0.162**      | 0.106  |
| Saturn | 0.053  | 0.192     | 0.144          | 0.092  |
| Ford   | **0.059** | 0.164  | 0.144          | **0.118** |
| Mazda  | 0.040  | 0.192     | 0.144          | 0.080  |

- **Étape 3:** b- déterminer la solution anti-idéale A'

    **A'= {0.040, 0.164, 0.144, 0.080}**

|        | Style  | Fiabilité | Carburant Eco. | Coût   |
|--------|--------|-----------|----------------|--------|
| Civic  | 0.046  | 0.244     | 0.162          | 0.106  |
| Saturn | 0.053  | 0.192     | **0.144**      | 0.092  |
| Ford   | 0.059  | **0.164** | **0.144**      | 0.118  |
| Mazda  | **0.040** | 0.192  | **0.144**      | **0.080** |

- **Étape 4:** a- distance idéale

    **A* = {0.059, 0.244, 0.162, 0.118}**

$S_i^* = \sqrt{[\sum_j (Vj* - Vij)^2]}$    pour chaque colonne j.

|        | Style | Fiabilité | Carburant Eco. | Coût |
|--------|-------|-----------|----------------|------|
| Civic  | $(0.046 - 0.059)^2$ | $(0.244 - 0.244)^2$ | $(0.162 - 0.162)^2$ | $(0.106 - 0.118)^2$ |
| Saturn | $(0.053 - 0.059)^2$ | $(0.192 - 0.244)^2$ | $(0.144 - 0.162)^2$ | $(0.092 - 0.118)^2$ |
| Ford   | $(0.059 - 0.059)^2$ | $(0.164 - 0.244)^2$ | $(0.144 - 0.162)^2$ | $(0.118 - 0.118)^2$ |
| Mazda  | $(0.040 - 0.059)^2$ | $(0.192 - 0.244)^2$ | $(0.144 - 0.162)^2$ | $(0.080 - 0.118)^2$ |

$$\sum_j (Vj* - Vij)^2 \qquad S_i^* = \sqrt{[\sum_j (Vj* - Vij)^2]}$$

| | | |
|---|---|---|
| Civic | 0.000313 | **0.01769** |
| Saturn | 0.0031316 | **0.05596** |
| Ford | 0.006724 | **0.082** |
| Mazda | 0.004829 | **0.06949** |

- **Étape 4:** b- distance anti-idéale

    **A' = {0.040, 0.164, 0.144, 0.08}**

    $S_i' = \sqrt{[\sum_j (Vj' - Vij)^2]}$   pour chaque colonne j.

| | Style | Fiabilité | Carburant Eco. | Coût |
|---|---|---|---|---|
| Civic | (0.046 – 0.040)² | (0.244 – 0.164)² | (0.162 – 0.144)² | (0.106 – 0.080)² |
| Saturn | (0.053 – 0.040)² | (0.192 – 0.164)² | (0.144 – 0.144)² | (0.092 – 0.080)² |
| Ford | (0.059 – 0.040)² | (0.164 – 0.164)² | (0.144 – 0.144)² | (0.118 – 0.080)² |
| Mazda | (0.040 – 0.040)² | (0.192 – 0.164)² | (0.144 – 0.144)² | (0.080 – 0.080)² |

$$\sum_j (Vj' - Vij)^2 \qquad S_i' = \sqrt{[\sum_j (Vj' - Vij)^2]}$$

| | | |
|---|---|---|
| Civic | 0.08046 | **0.283** |
| Saturn | 0.001097 | **0.0331** |
| Ford | 0.001801 | **0.0000032** |
| Mazda | 0.000784 | **0.00000061** |

- **Étape 5:** Calculer la proximité relative à la solution idéale.

    $C_i^* = S_i' / (S_i^* + S_i')$

|        | $S'_i / (S_i^* + S'_i)$ | $C_i^*$ |
|--------|-------------------------|---------|
| Civic  | 0.283/0.30069           | **0.94116** → la meilleure solution |
| Saturn | 0.0331/0.08906          | 0.37165 |
| Ford   | 0.0000032/0.0820032     | 0.000039 |
| Mazda  | 0.00000061 /0.06949     | **0.0000087** → la mauvaise solution |

## I.5  JFreeChart

JFreeChart est une librairie open source sous Licence LGPL (Lesser General Public Licence) permettant de créer rapidement des graphiques pour vos statistiques**.**
Il est possible de choisir le type de rendu des graphes:

- Composants Swing.
- Images (PNG ou JPEG).
- Fichiers d'images vectorielles PDF, EPS et SVG.

## I.6  Conclusion

Dans ce chapitre, nous avons présenté des notions importantes et domaines d'applications concernant la radio cognitive, les systèmes multi-agents et la décision multicritère.

Nous avons aussi introduit les outils JADE et JFreeChart que nous les utilisons pour la réalisation de notre application dans le chapitre suivant.

# CHAPITRE II : Evaluation des résultats obtenus à l'aide de JADE

Chapitre II: Evaluation des résultats obtenus à l'aide de JADE

## II.1  Introduction

Dans le cadre de ce travail, notre travail consiste à aider les SU à choisir les PU qui répondent aux mieux à leurs exigences en termes de qualité de service. Pour réaliser cette tâche, nous avons utilisé une des méthodes de la décision multicritère qui est TOPSIS en faisant une gestion par coalitions. Nous avons créé des coalitions CPU et CSU pour jouer le rôle de coordinateurs pour l'ensemble des PU et SU respectivement. Nous avons aussi appliqué la technique d'agrégation de messages dans le but de minimiser le nombre de messages échangés dans le réseau.

## II.2  Topologie du réseau utilisé

Dans ce qui suit, nous allons utiliser un réseau ad-hoc car il est moins coûteux (sans infrastructure) et s'adapte facilement à n'importe quel endroit (champ de guerre, lieu des catastrophes naturelles,…).

On suppose que les PU sont répartis sur des zones géographiques bien précises sans changer de position durant la négociation. Selon un critère de rapprochement géographique, chaque agent de coalition CPU regroupe les PU qui se trouvent dans sa zone géographique.
On fera un schéma identique pour les CSU vis-à-vis des SU.
A noté que les CPU et CSU jouent le rôle d'intermédiaires entre les SU et les PU.

Dans la figure ci-dessous, on donne un exemple de négociation illustré par la topologie suivante du réseau : 15 PU réparties uniformément en 5 CPU et 15 SU réparties (non nécessairement uniforme) en 3 CSU.

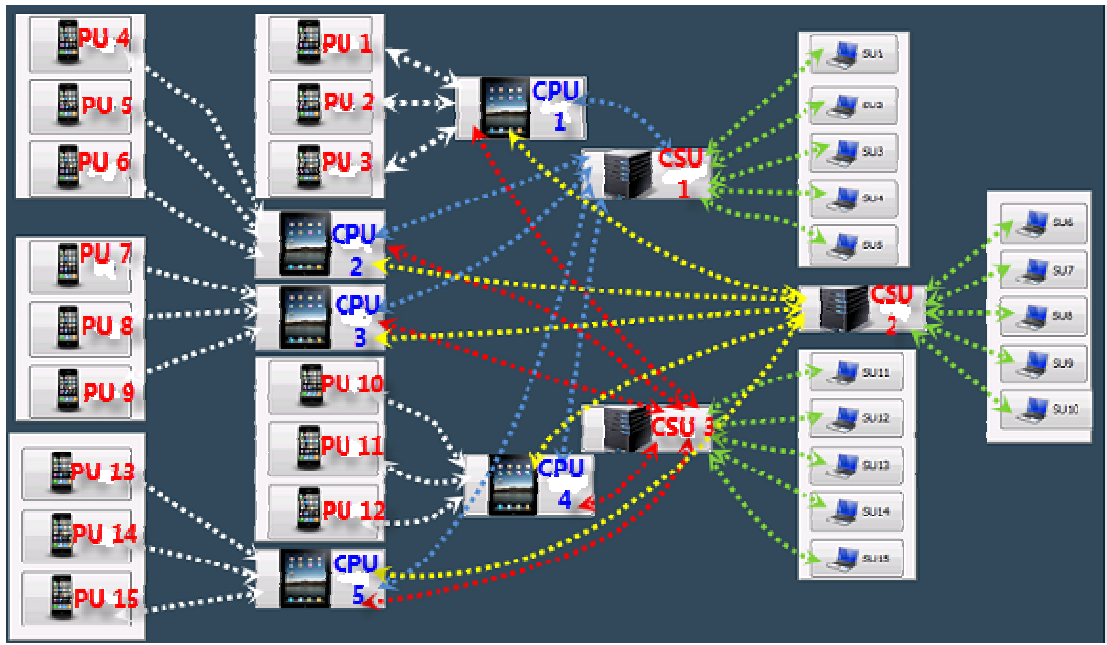

**Figure II.1:** Illustre la topologie du réseau dont nous nous sommes servi

## II.3  Présentation du fonctionnement

Notre démarche consiste à créer un modèle typique de négociation : « de plusieurs à plusieurs ».

Mais comme le nombre d'intervenant, les PU et les SU est important donc la négociation devient fastidieuse. Nous avons alors décidé d'implémenter des intermédiaires (des CPU et des CSU) répartis sur des zones géographiques différentes. Chaque CPU détient, en permanence et en temps réel, l'ensemble des informations relatives à l'ensemble des PU présents à l'intérieur de sa sphère. Concernant les CSU, chacun d'entre eux détient aussi l'ensemble des informations relatives aux SU qui ont demandés d'allouer des canaux présents dans sa zone géographique. Les SU laissent la mission de négociation avec les CPU au CSU. Ce dernier utilise l'agrégation de messages afin d'entamer la phase de négociation. Dans notre cas, les CSU et les CPU négocient leur accord sur une base de multiples critères tels que le prix, le nombre de canaux, ainsi que le temps d'allocation. Cette multitude de paramètres nous conduit à introduire une méthode adéquate qui est TOPSIS.

La figure II.2 nous montre qu'une fois le SU ou les SU cherche (ent) à allouer des canaux et après un temps bien déterminé le (les) CSU (s) va (vont) initier des négociations avec les CPU qui possèdent l'ensemble des informations sur les PU présents. Chaque CPU applique l'algorithme TOPSIS pour choisir l'offre le mieux adapté au besoin du CSU, puis envoie à ce dernier l'offre désigné. Quant au CSU, il aura à réceptionner à son tour les 5 offres sélectionnées en appliquant l'algorithme TOPSIS pour aboutir à un (des) choix optimal (optimaux).

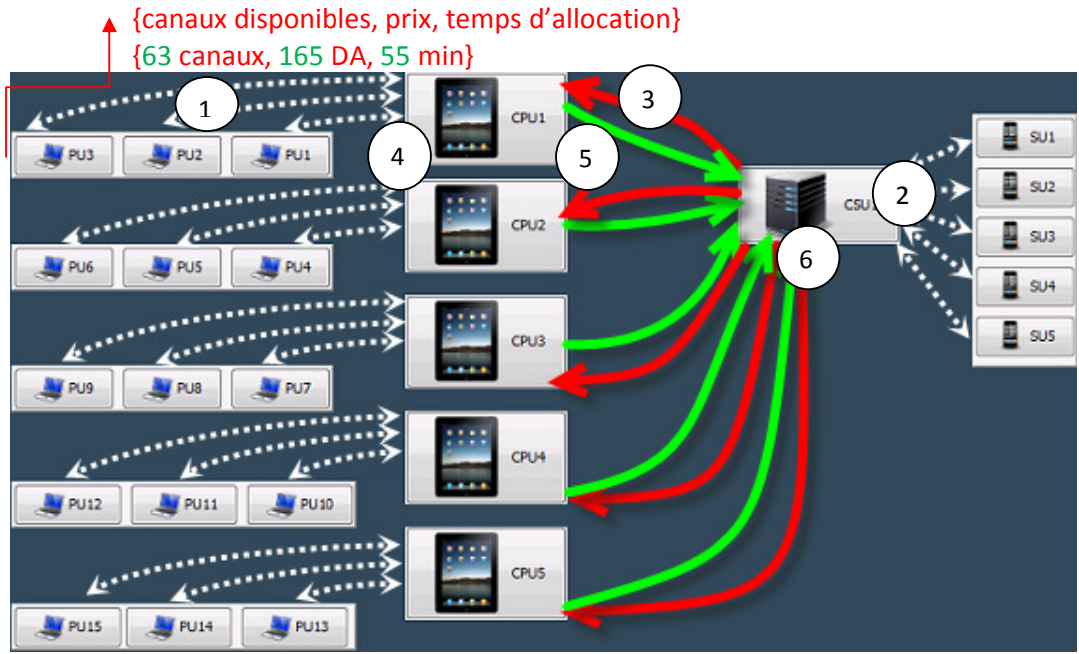

**Figure II.2:** La négociation entre les SU et PU à l'aide des intermédiaires CPU et CSU

① Chaque PU envoie ses paramètres {canaux disponibles, prix, temps d'allocation} au CPU le plus proche géographique.

② Les SU cherchent d'allouer des canaux libres.

③ Le CSU contacte les CPU pour leur signaler le besoin des SU qui sont dans sa zone.

④ Chaque CPU sélectionne la meilleure offre parmi celles présentes dans sa liste en appliquant l'algorithme TOPSIS.

⑤ Chaque CPU renvoie au CSU l'offre sélectionnée.

⑥ Les CSU reçoivent toutes les réponses des CPU. Grâce à l'algorithme TOPSIS ils obtiennent les offres optimales et les renvoient aux SU.

## II.4 Implémentation de l'application

Cette partie est consacrée au fonctionnement de notre application, aux résultats obtenus et aux comparaisons effectuées avec d'autres solutions déjà implémentées auparavant. La réalisation de cette application a été faite sous Netbeans version 8.0.1 en se basant sur JADE version 4.3.3 avec l'utilisation de l'outil JFreeChart version 1.0.19 pour créer des graphiques et des diagrammes de très bonne qualité et le développement des interfaces a été réalisé à l'aide des bibliothèques SWING et AWT.

### II.4.1 Présentation de l'application

Notre application se compose de 15 SU, 5 CSU, 5 CPU et de 15 PU. Nous avons opté qu'un CPU regroupe trois PU et qu'un CSU peut regrouper 5 SU. Ainsi chaque CPU renfermera les informations sur les paramètres relatifs à ces trois PU se trouvant dans sa zone géographique.

Chaque PU détiendra trois paramètres (prix, temps d'allocation, nombre de canaux). Au lancement de la simulation, chaque PU envoie à son CPU correspondant les informations sur ses paramètres (critères). Quant au CPU, il chargera les informations reçus dans une table de hachage de type HashMap dont les clés sont les noms des PU qui pointent vers des tables qui contiennent les noms des PU et les 3 critères cités auparavant. Un exemple de HashMap utilisée coté CPU est indiqué dans la figure II.3

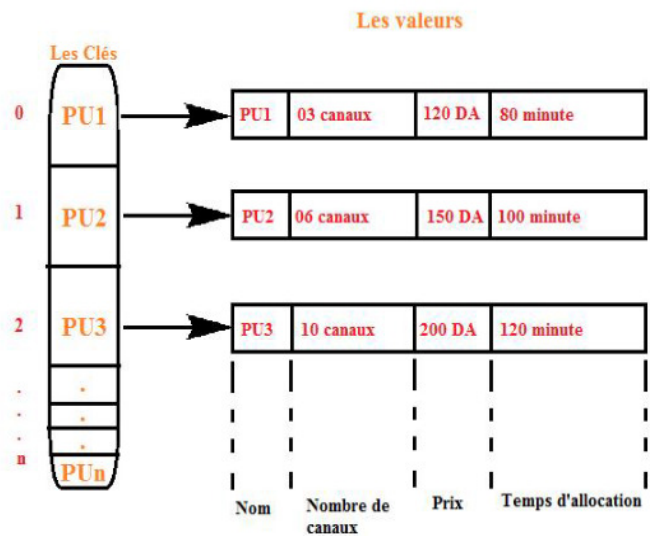

**Figure II.3:** Table de hachage implémentée coté CPU

La table de hachage est entretenue en temps réel en cas de changement de paramètre chez un PU.

Cette limitation dans le nombre de CPU et de PU a été faite afin d'éviter que le temps de la simulation ne devienne long et fastidieux.

### II.4.2   Interactions entre les CSU et les CPU

La méthode TOPSIS est applicable dans n'importe quelle étude contenant une multitude de critères en tenant compte des poids associés aux critères. Pendant l'implémentation de cette application, nous avons opté pour les poids suivants {0.2, 0.5, 0.3} correspondant respectivement aux trois critères {nombre de canaux, prix, temps d'allocation}. L'objectif est d'obtenir une solution idéale qui est de maximiser le nombre de canaux et le temps d'allocation et minimiser le prix. A noter que la solution anti-idéale est de réduire le nombre de canaux et le temps d'allocation, et de maximiser le prix.

L'avantage de l'agrégation de messages pendant l'interaction entre les CPU et les CSU est de minimiser le nombre de messages et ainsi éviter la saturation du réseau.
Plus la capacité de CSU est grande et le nombre de CPU et CSU est important, plus le rôle de l'agrégation devient très utile et nécessaire.

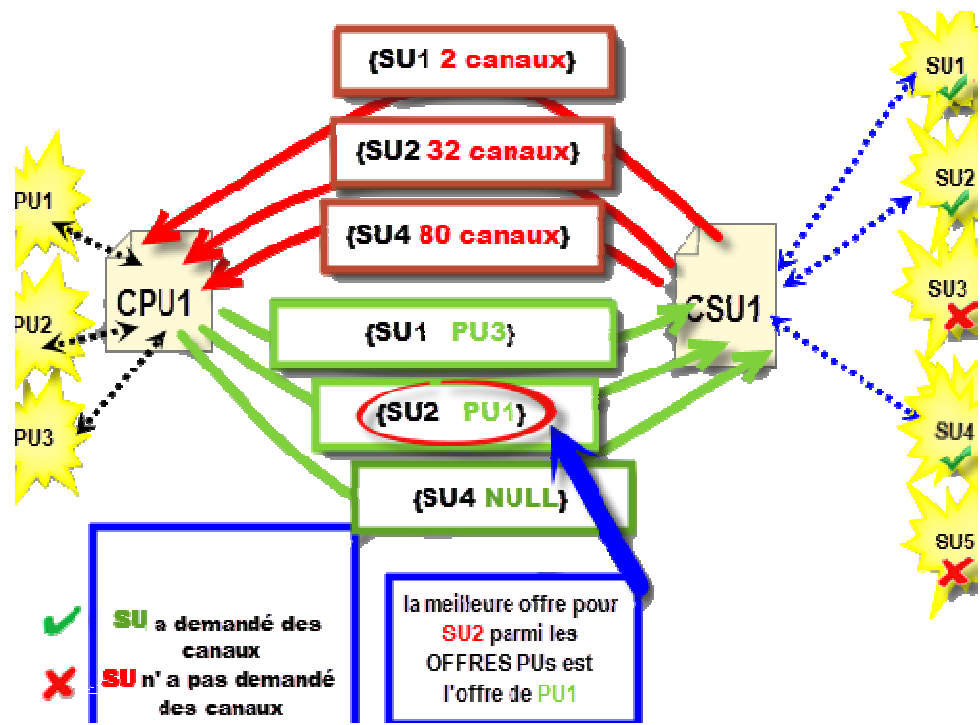

**Figure II.4:** Un exemple de communication entre CPU et CSU sans agrégation

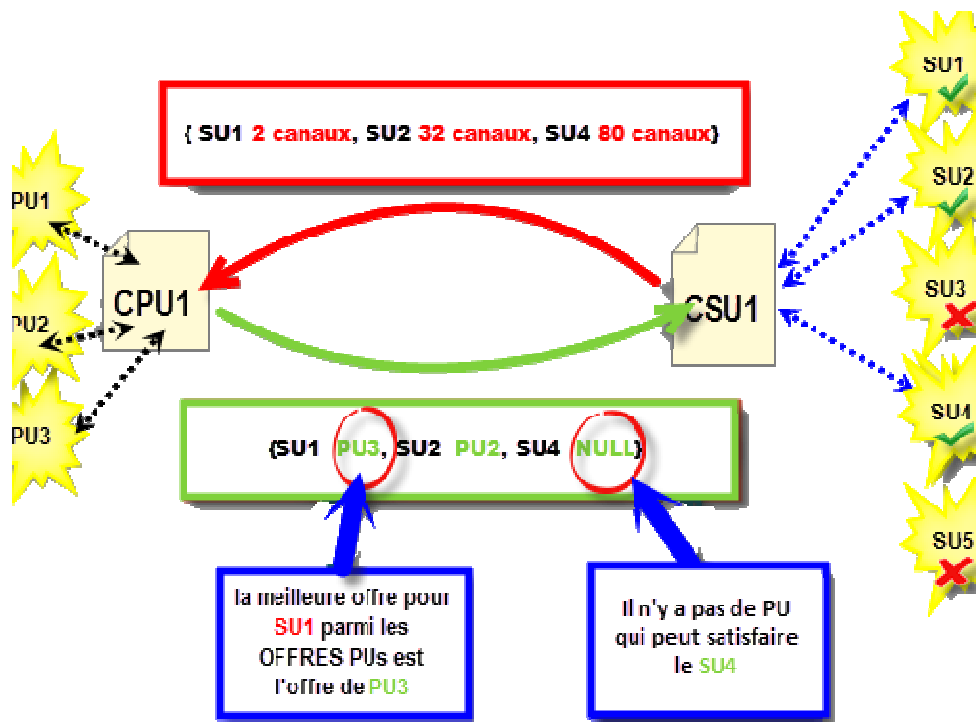

**Figure II.5:** Un exemple de communication entre CPU et CSU avec agrégation

## II.4.3 Etude comparative

Dans cette sous-section, nous donnons les quatre études comparatives suivantes:

i. **Temps de réponse en fonction du nombre de SU dans le CSU**

La nécessité de recevoir une réponse rapide à sa demande, en la présence de plusieurs offres, pousse le SU à considérer le temps de réponse comme étant un paramètre important.

Pour étudier l'impact de la taille du CSU (en termes de nombre de SU) sur le temps de réponse, nous avons considéré la topologie suivante : 15 PU réparties sur 5 CPU d'une manière régulière et 1 seul CSU contenant dans chacun des cas traités: 1, 2, 3, 4, 5 et 10 SU. La figure II.6 représente le résultat obtenu.

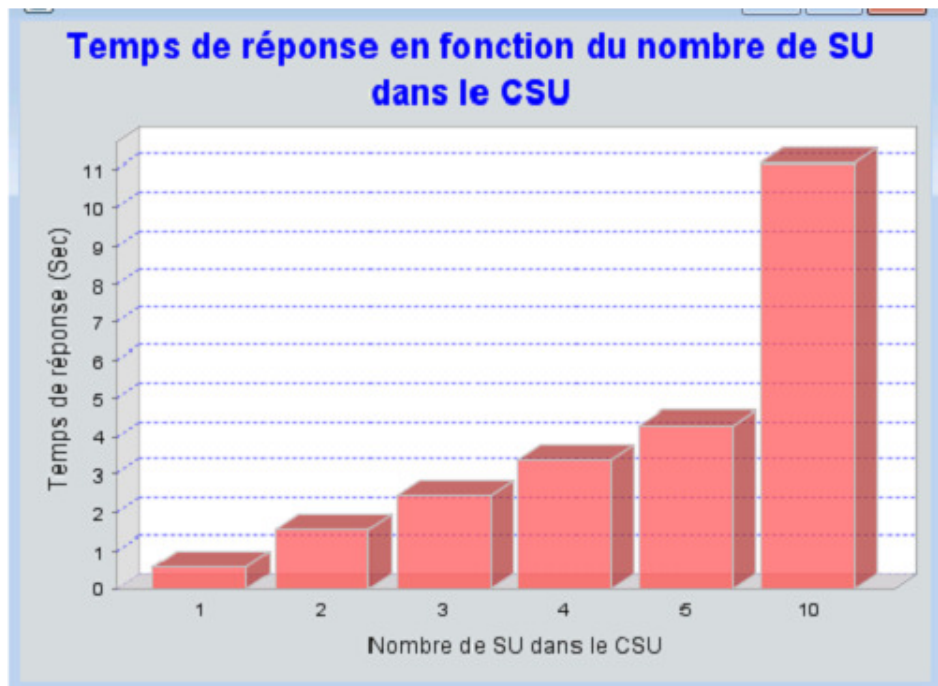

**Figure II.6:** Impact de la capacité du CSU sur le temps de réponse

**Conclusion 1 :** On remarque que le temps de réponse dépend de la capacité de chaque CSU, c'est-à-dire que plus le nombre de SU dans le CSU est important plus le temps de réponse est aussi important. Le CSU utilise dans ce cas l'agrégation de messages donc il est obligé d'attendre les demandent provenant de l'ensemble des SU. Plus le nombre de SU est important plus le temps d'attente côte CSU pour recevoir toutes les demandent des SU est aussi important.

### ii. Temps de réponse par rapport au nombre de CSU

Dans ce qui suit, nous allons mesurer l'impact du nombre de CSU sur le temps de réponse. Nous avons fixé le nombre de SU à 10 et nous avons considéré trois configurations différentes: 5 CSU dont chacun contient 2 SU, 2 CSU dont chacun contient 5 SU et enfin 1 CSU qui contient les 10 SU.

Pour les trois configurations précédentes, les temps de réponse sont respectivement : 3.5 sec, 6.21 sec et 12.06 sec comme indiquer dans la figure suivante**.**

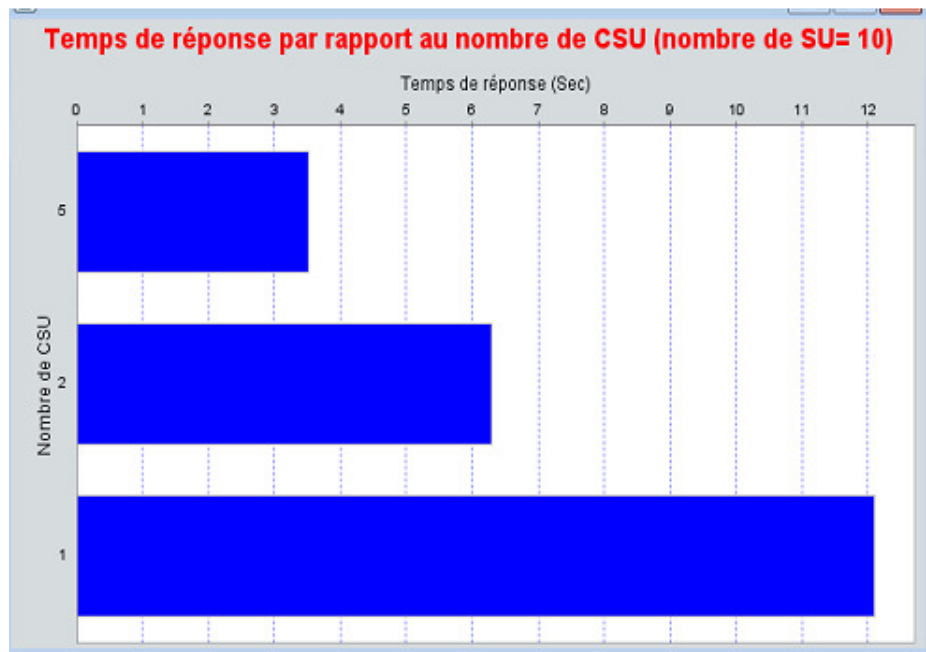

**Figure II.7:** Temps de réponse par rapport au nombre de CSU

**Conclusion 2 :** Le nombre de CSU a son impact sur le temps de réponse. Le mieux est de répartir les SU sur le maximum de CSU afin de réduire le temps de réponse et bénéficier de l'exécution parallèle du traitement dans les différents CSU.

### iii. Nombre de messages en fonction du nombre de CSU

Dans ce qui suit, nous allons considérer 1000 SU qui seront répartis selon plusieurs configurations sur plusieurs CSU comme suit: 500 CSU dont chacun contient 2 SU, 100 CSU dont chacun contient 10 SU, 40 CSU dont chacun contient 25 SU et enfin 1 CSU qui contient les 1000 SU.

L'objectif est d'étudier l'impact du nombre de CSU sur le nombre de messages échangés dans le réseau comme indiquer dans la figure suivante.

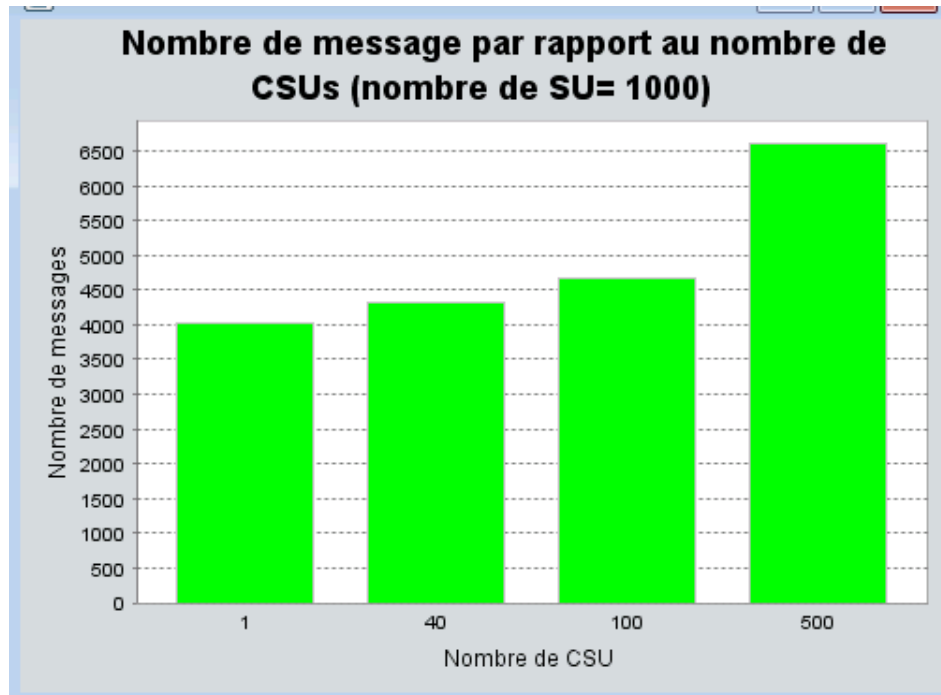

**Figure II.8:** Impact du nombre de CSU sur le nombre de messages

**Conclusion 3 :** Le nombre de messages augmente lorsque le nombre de CSU augmente.

iv. **Nombre de messages dans trois topologies différentes : sans coalitions, avec CPU et sans CSU, avec CPU et CSU**

Nous allons étudier l'influence de la topologie du réseau sur le nombre de messages échanges. Pour cela, nous avons considéré les trois topologies suivantes :

- ✓ **Topologie 1 (sans coalitions):** 15 PU, nombre variable de SU, 0 CPU et 0 CSU.
- ✓ **Topologie 2 (avec CPU et sans CSU):** 15 PU, nombre variable de SU, 5 CPU de taille homogène et 0 CSU.
- ✓ **Topologie 3 (avec CPU et CSU):** 15 PU, 5 CPU, nombre variable de SU et de CSU (de même taille: 5 SU dans le CSU) .

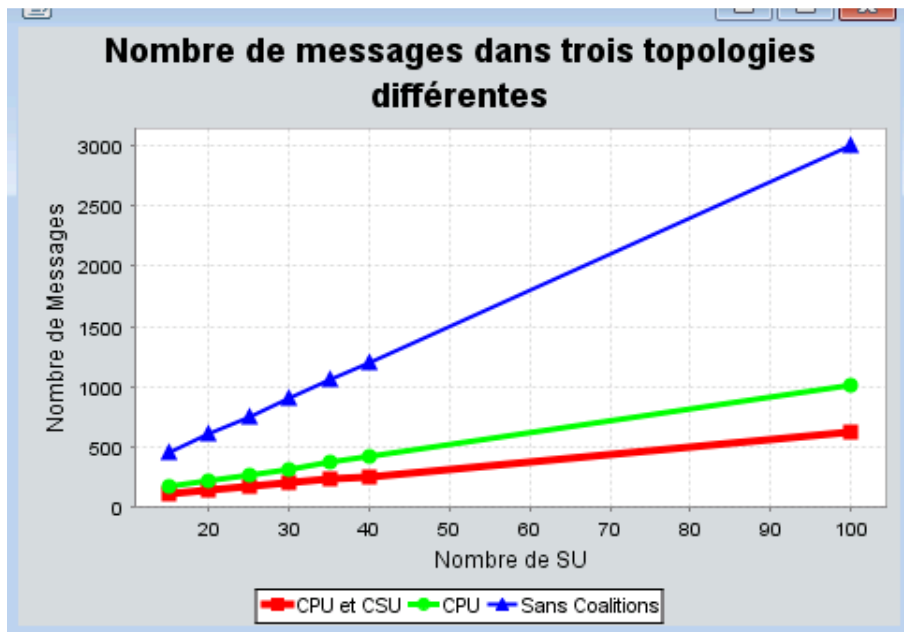

**Figure II.9:** Comparaison en termes de nombre de messages dans trois topologies différentes

Nous distinguons qu'à chaque fois le nombre des SU augmente, le nombre de messages augmente à son tour puisque les interactions s'amplifient. En ce qui concerne la comparaison dans les trois cas (sans coalition, avec CPU et sans CSU, avec CPU et CSU), l'absence des coalitions pousse les SU à interroger les PU un par un. Cependant, l'implémentation des CPU réduit le nombre de messages car chaque CPU regroupe plusieurs PU et de ce fait les SU interrogent un nombre restreint d'agents. D'où la diminution du nombre de messages par rapport au cas classique (sans coalitions) mais reste assez grand par rapport à notre cas: la présence des CPU, des CSU et l'utilisation de l'agrégation des messages.

**Conclusion 4 :** L'utilisation des coalitions et la technique d'agrégation permettent de diminuer le trafic des messages échangés entre les agents, car chaque CPU et CSU regroupent plusieurs PU et SU. Par conséquent, la négociation se fait uniquement entre les CPU et les CSU.

## II.5  Conclusion

Dans ce chapitre, nous avons présenté le fonctionnement de notre application en détail, la topologie du réseau utilisé et le modèle de négociation (de plusieurs SU à plusieurs PU). Nous avons montré l'importance de la gestion par coalitions dans ce contexte en se basant sur les techniques de décisions multicritère comme TOPSIS mais aussi sur l'agrégation de messages. Les résultats obtenus sont intéressants car ils permettent de réduire le nombre de messages échangés dans le réseau et aussi le temps de réponse pour l'utilisateur de la RC.

# CONCLUSION GÉNÉRALE

# Conclusion générale

Dans ce projet de fin d'études, on s'est rendu compte de l'importance des réseaux de radio cognitive qui utilisent le spectre radio d'une manière opportuniste et de l'intérêt de la méthode TOPSIS dans les négociations afin de satisfaire au mieux les besoins des SU en fonction de plusieurs critères.

L'utilisation de plusieurs CPU et de plusieurs CSU dans ce contexte ainsi que l'agrégation des messages côté CSU permettent de réduire considérablement le temps de réponse coté SU mais aussi le nombre de messages échangés dans le réseau et ainsi éviter la surcharge de ce dernier.

Pour les perspectives, il serait intéressant de considérer une topologie dynamique de réseau où les nœuds sont mobiles dans le cadre par exemple d'un réseau de type MANET (Mobile Ad-hoc Networks). On propose également de généraliser notre étude au cas où le nombre de CPU et de CSU devient très grand en définissant de nouvelle coalition pour les coalitions déjà existantes (des CCPU et des CCSU).

# RÉFÉRENCES BIBLIOGRAPHIQUES

# Références bibliographiques

# Références bibliographiques

# LISTE DES FIGURES

# LISTE DES TABLEAUX

# LISTE DES ABRÉVIATIONS

| Acronym | Signification |
|---|---|
| RC | **R**adio **C**ognitive |
| RF | **R**adio **F**reqency |
| SDR | **S**oftware **D**efined **R**adio |
| AACR | **A**daptative **A**ware **C**ognitive **R**adio |
| OSI | **O**pen **S**ystems **I**nterconnection |
| SP | **S**ensory **P**erception |
| RRC | **R**éseaux de **R**adio **C**ognitive |
| URC | **U**tilisateur à **R**adio **C**ognitive |
| UP | **U**tilisateur **P**rimaire |
| SMA | **S**ystème **Multi-agents** |
| IAD | **I**ntelligence **A**rtificielle **D**istrubuée |
| PU | **P**rimary **U**ser |
| SU | **S**econdary **U**ser |
| CPU | **C**oalition **P**rimary **U**ser |
| JADE | **J**ava **A**gent **De**velopment Framework |
| LAMSADE | **L**aboratoire d'**A**nalyse et **M**odélisation de **S**ystèmes pour l'**A**ide à la **DE**cision |
| FIPA | **F**oundation for **I**ntelligent **P**hysical **A**gents |
| RMI | **R**emote **M**ethod **I**nvocation |
| DF | **D**irector **F**acilitor |
| ACC | **A**gent **C**ommunication **C**hannel |
| AMS | **A**gent **M**anagement **S**ystem |
| ACL | **A**gent **C**ommunication **L**anguage |
| LGPL | **L**esser **G**eneral **P**ublic **L**icence |
| ELECTRE | **EL**imination **E**t **C**hoix **T**raduisant la **RE**alité |
| AHP | **A**nalytic **H**ierarchy **P**rocess |
| WPM | **W**eight **P**roduct **M**ethod |
| WSM | **W**eight **S**um **M**ethod |
| TOPSIS | **T**echnique for **O**rder **P**reference by **S**imilarity to **I**deal **S**olution |

# ANNEXE

## La plateforme JADE

### 1) Présentation

JADE est développée par le laboratoire Télécom Italia (http://jade.tilab.com/) et vise à simplifier la mise en œuvre de systèmes multi-agents distribués à travers un middleware qui se conforme aux spécifications de FIPA (Fountion for Intelligent Physical Agents) et à travers un ensemble d'outils qui prennent en charge les phases de débogage et de déploiement [2]. La plateforme peut être distribuée sur plusieurs ordinateurs et la configuration peut être contrôlée à partir d'une interface graphique à distance. La configuration peut même être changée au moment de l'exécution par des agents mobiles d'une machine à une autre en cas de besoin. L'implémentation d'un agent se fait par l'intermédiaire du langage Java, tandis que la communication s'effectue à l'aide de RMI (Remote Method Invocation).

Ces descriptions succinctes permettent de voir que certaines plateformes ont déjà été évaluées, voire conçu, pour des systèmes parallèles de type cluster alors que d'autres sont plus orientées sur les systèmes distribués, moins fortement couplés, de type réseaux de stations de travail.

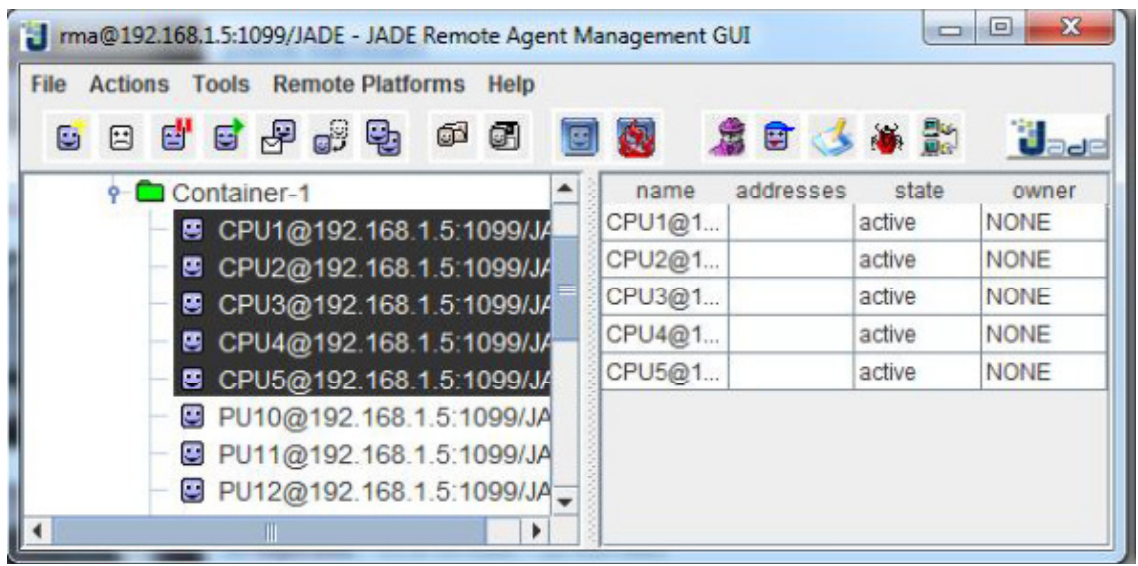

**Figure A. 1:** L'interface graphique de JADE

- ➢ JADE possède trois modules principaux (nécessaires aux normes FIPA) :
    - ✓ DF « Directory Facilitator » fournit un service de « pages jaunes» à la plate-forme.
    - ✓ ACC «Agent Communication Channel » gère la communication entre les agents.

- ✓ AMS « Agent Management System » supervise l'enregistrement des agents, leur authentification, leur accès et l'utilisation du système.

➢ Ces trois modules sont activés à chaque démarrage de la plate-forme.

**2) Agents prédéfinis**

JADE propose différents agents prédéfinis :

a. **DUMMY AGENT**
- ✓ Permet l'envoi des messages aux agents.
- ✓ Utile lors de la conception afin de vérifier la réaction d'un agent à la réception d'un message.

b. **SNIFFER AGENT**
- ✓ Permet de visualiser l'enchaînement des messages entre les agents.
- ✓ Représentation graphique de l'échange des messages.
- ✓ Les agents à sniffer peuvent être sélectionnés.

c. **INTROSPECTOR AGENT**
- ✓ Permet de débugger le comportement d'un agent.
- ✓ Permet de savoir quels comportements d'un agent sont exécutés, dans quel ordre et quand.

**3) Installation :** Pour manipuler JADE sur la ligne de commande, vous aurez besoin d'ajouter la variable d'environnement classpath du **fichier jade.jar** téléchargeable sur : **http://jade.tilab.com/jade-4-3-3-wade-3-4-and-amuse-1-0-have-been-released/**

**4) Démarrage**

- ✓ Lancer Jade avec la ligne de commandes :    java jade.Boot
- ✓ Lancer Jade et la GUI :    java jade.Boot –gui
- ✓ Lancer un agent au démarrage :

    java jade.Boot –gui <nom de l'agent>:<classe de l'agent>

- ✓ Lancer un agent avec des paramètres :

    java jade.Boot –gui <nom de l'agent>:<classe agent>    (<Param>).

- ✓ Lancer un autre container dans une autre machine :

java jade.Boot –container –host Main_IP_Adress

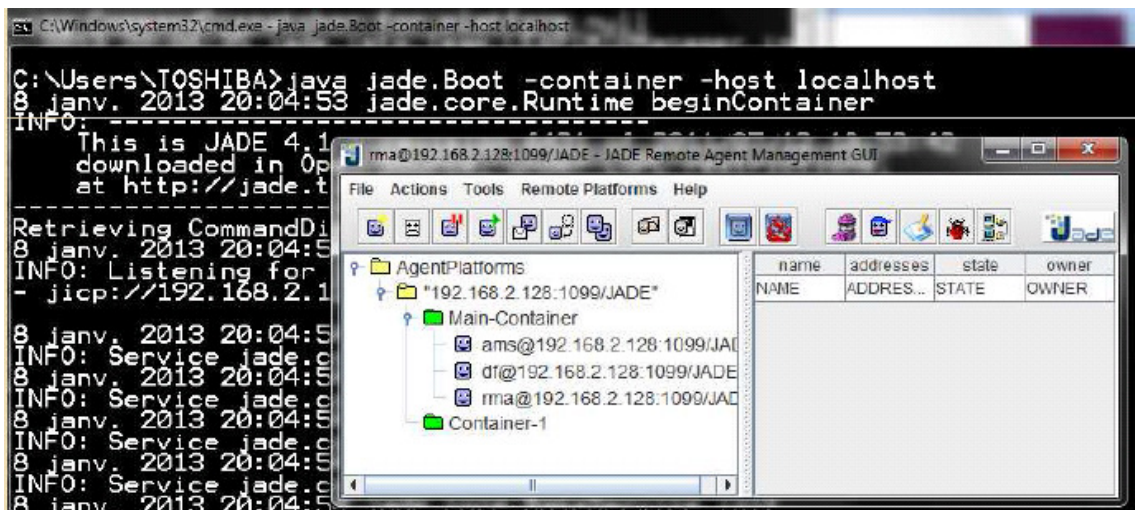

**Figure A.2:** Lancement un autre container « Container-1 »

5) **Création d'un agent**

✓ Etendre la classe jade.core.Agent

    import jade.core.agent;

    public class monAgent extends Agent …;

✓ Chaque agent est identifie par un AID

    Methode getAID() //pour récupérer l'AID

✓ Dans la méthode setup() (Obligatoire)

- Enregistrer les langages de contenu.
- Enregistrer les Ontologies.
- Enregistrer les Services auprès du DF.
- Démarrer les Comportements (behaviors).

6) **Identification des agents**

✓ Le nom d'un agent :

    <nom-agent>@<nom-plate-forme>   Doit être globalement unique

✓ Plate-forme par défaut :

  <main-host>:<main-port>/JADE

✓ Nom de la plate-forme défini avec –name.